\documentclass{pasj00}
\draft

\begin{document}
\SetRunningHead{H. Enoguchi et al.}{Kes\,27: A Typical Mixed-Morphology SNR}
\Received{2001/10/24} 
\Accepted{2002/02/20} 

\title{Observation of Kes 27: \\
A Typical Mixed-Morphology SNR}

\author{Hideyuki \textsc{Enoguchi},
Hiroshi \textsc{Tsunemi},
Emi \textsc{Miyata},
and
Kumi \textsc{Yoshita}
}
\affil{Department of Earth and Space Science,
        Graduate School of Science, Osaka University,\\
        1-1 Machikaneyama, Toyonaka, Osaka 560-0043}
\email{enoguchi@ess.sci.osaka-u.ac.jp, tsunemi@ess.sci.osaka-u.ac.jp,\\
       miyata@ess.sci.osaka-u.ac.jp, kyoshita@ess.sci.osaka-u.ac.jp}

\KeyWords{ISM: abundances --- ISM: individual (Kes\,27) --- supernova
remnants --- X-rays: ISM)}

\maketitle

\begin{abstract}

 We report here on the observation of Kes\,27, a proto-typical
 mixed-morphology SNR, using ASCA.  It clearly shows a 
 filled-center structure in the X-ray region while a shell structure in
 the radio region.  There are two radio bright regions: one is in the
 center, while the 
 other is in the east rim.  The X-ray intensity peak coincides well 
 with
 the radio bright region at the center.  The X-ray spectrum was
 well-fitted by a collisional ionization equilibrium model with solar
 abundances. Taking into account the ionization parameter
 ($>$10$^{12}~$cm$^{-3}$s) and the plasma density
 (0.39$^{+0.03}_{-0.02}~$cm$^{-3}$), we found that the age of the SNR is
 longer than $8 \times 10^{4}$ yr. The hardness ratio map indicates
 that the inner region shows a harder spectrum than that in the outer
 region, which does not come from the heavier interstellar absorption
 feature, but from the higher temperature. There is a temperature
 gradient from the innner region (0.84$\pm$0.08~keV) toward outer 
 region (0.59$^{+0.04}_{-0.06}$~keV), indicating that the thermal conduction
 does not play an important role.

\end{abstract}

\section{Introduction}

There are two types of Supernova remnants (SNRs) in the X-ray region from
a morphological point of view: a shell structure and a filled-center
structure.  Some SNRs in each type contain a point source at its center.
Therefore, the point source at its center does not always affect the
morphology of the SNR.  The shell structure is generated as the result of
a blast wave propagating inside the interstellar matter (ISM), while
the origin of the filled-center structure has not yet been established. A
cloud evaporation model can explain the filled-center structure (White,
Long\ 1991). The clouds in the interstellar space will gradually
evaporate after passage of the blast wave, which enhances the
brightness of the central part of the SNR, resulting in the filled-center
structure.  Many of them show evidence of the interactions with
molecular clouds, and have been studied from a theoretical point of 
view (Chevalier\ 1999). Another explanation is a radiative
phase model (Cox\ 1972). When the density at the shell region increases,
the radiative instability proceeds and reduces the temperature of the
shell region so that it becomes X-ray dim in the shell region.

Rho and Petre (1998) proposed a
new group of SNR, ``mixed-morphology (MM) SNR'', among the filled-center
SNRs based on a ROSAT observation.  They are characterized as having
1) a shell structure in the radio region, 2) a filled-center structure
in the X-ray region, 3)an absence of a compact source in its center, and
4) thin thermal emission in the X-ray region showing a solar or sub-solar
abundance.  The forth criterion is considered to mean that the X-ray
emitting plasma is not contaminated by the ejecta.  They selected
7\,SNRs belonging to this group as proto-typical MM SNRs: W\,28, W\,44,
3C\,400.2, Kes\,27, MSH\,11-61A,
3C\,391 and CTB\,1.  They also claimed 7 other SNRs as probably belonging
to this group. 

In young SNRs, like Cassiopeia-A, there is a large inhomogeneity in
metal abundance (Hwang et al.\ 2000).  Even in a middle-aged
SNR, like the Cygnus Loop, the abundance in the shell region is quite
different from that in its center (Miyata et al.\ 1998, 1999), whereas
the absolute intensity in the center is quite weak.  Although these two
SNRs belong to the shell structure, it will take long time before the
convection of the ejecta with the ISM is completed.  Therefore, the
uniform abundance in the SNR may become an important clue to form a new 
class of SNRs.

Among the MM SNRs, Yoshita et al.\ (2001) reported on the X-ray structure of
3C\,400.2 using the ASCA satellite, which has better energy resolution
than that of ROSAT.  They found no spectral variation across the
SNR, with a possible exception of the abundance of Fe.  Due to an
elongation in the radio image, they studied a possible interaction of two
SNRs.  They concluded that it was generated from a single SNR, rather than two.

The SNR, Kes\,27 (G327.4+0.4), is classified as a proto-typical MM type
SNR.  In the radio wavelength, it shows a shell structure with a
diameter of $\sim\,20^{\prime}$ with a slight complexity, a typical
shell with an arm in the northwest.  It has a spectral index of $\alpha
= 0.6\pm 0.05$ where ${\rm S}_{\nu}\sim\nu^{-\alpha}$ (Milne et al.\
1989), which is a typical value of shell-type SNRs in the radio region.
No optical emission has been detected from Kes~27 (van den Bergh\ 1978; 
Kirshner, Winkler\ 1979).

The X-ray observation of Kes\,27 was initially motivated by the fact
that the remnant was located within the error circle of the COS\,B
unidentified $\gamma$-ray source, CG327--0 (Hermsen et al.\ 1977).  Lamb 
and Markert (1981) observed Kes\,27 with the Einstein IPC, and found
that the X-ray emission was centrally peaked.  An IPC image was also
given by Seward (1990), which clearly showed clumpy X-ray emission.
Seward et al.\ (1996) reported using ROSAT data that there
were several unresolved point-like sources as well as diffuse
emission.  It showed not only emission from inside, but also that
from the bright eastern shell, which coincides with the bright radio
emission.  Spectral studies of the diffuse emission showed little
difference between the central region and the eastern rim.

We report here on an observation of Kes\,27 using the ASCA satellite
to study its X-ray structure, particularly its spectral  variation.

 \section{Observations}

The ASCA observation of Kes\,27 was performed on 1994 August
21$-$22.  We retrieved these data from DARTS Astrophysical Database at
the ISAS PLAIN center.  The SISs (Yamashita et al.\ 1997) were operated
in a combination of the 2-CCD BRIGHT mode and of the 4-CCD FAINT mode.  Some
data were obtained in the 2-CCD mode, while others were obtained in the
4-CCD mode.  Since the radio shell of Kes\,27 has the diameter of $\sim
20^{\prime}$, most of the remnant can be covered by the field of view
(fov) of the SIS in the 4-CCD mode, which is a square of $22^{\prime}
\times 22^{\prime}$. Since the major part of the observation was done in
the 4-CCD mode, we only selected the 4-CCD FAINT mode data. We excluded
all of the data taken at an elevation angle from the Earth rim below
$5^{\circ}$ from the night Earth rim and $40^{\circ}$ (SIS~0) or
$20^{\circ}$ (SIS~1) from the day Earth rim, a geomagnetic cutoff
rigidity lower than 6\,GV, and the region of the South Atlantic Anomaly.
After screening the above criteria, we
further removed the time region of a sudden change of the corner pixels
of X-ray events.  We then removed the hot and flickering pixels and
corrected CTI, DFE and Echo effects (Dotani et al.\ 1995) from our data
set. The exposure times after the
screening were 9\,ks for SIS~0, and 11\,ks for SIS~1.  The GISs
were operated in the PH mode with the standard bit assignments
(Makishima et al.\ 1996).  The GIS data were also screened in a
different way.  We excluded all of the data taken at an elevation angle from
the Earth rim below $5^{\circ}$, a geomagnetic cutoff rigidity lower
than 6\,GV, and the region of the South Atlantic Anomaly.
The exposure times after the screening were 13\,ks for GIS~2 and GIS~3,
respectively.

\section{Analysis and Results}\label{sec:kes27_ana}

\subsection{GIS Images}
\begin{figure}
\FigureFile(85mm,85mm){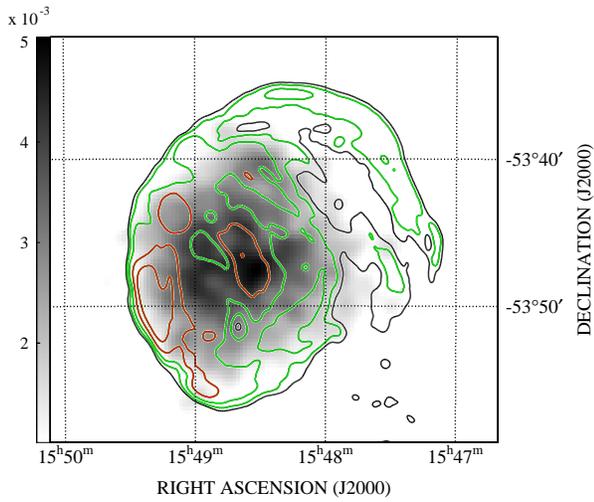}
\caption{GIS image in the 0.7$-$10\,keV band overlaid on the MOST
 contours.  The MOST data were retrieved from the on-line version of the
 MOST Supernova Remnant Catalogue at
 $\langle$http://www.physics.usyd.edu.au/astrop/wg96cat/$\rangle$
 (Whiteoak, Green\ 1996). Radio contours are logalismically spaced (Red,
 Green, and Black): red is the strongest, followed by green and black.} 
\label{fig:kes27_gismost}
\end{figure}

We subtracted the non-X-ray background and the cosmic X-ray background
from the data.  We then made a correction for vignetting.
Figure~\ref{fig:kes27_gismost} shows the GIS image of Kes\,27 in the
0.7--10\,keV energy band. The overlaid contours are the radio map
reproduced from the Molonglo Observatory Synthesis Telescope (MOST)
observation (Milne et al.\ 1985, 1989).  It clearly shows the centrally
peaked X-ray emission confined by the radio shell.  In the radio map,
there are two bright regions drawn in red contour: one is the east rim
and the other is the center.  Therefore, the radio-bright rim in the
east is relatively dim in X-rays, while the radio-bright center is
bright in X-rays.

\begin{figure}
\FigureFile(85mm,85mm){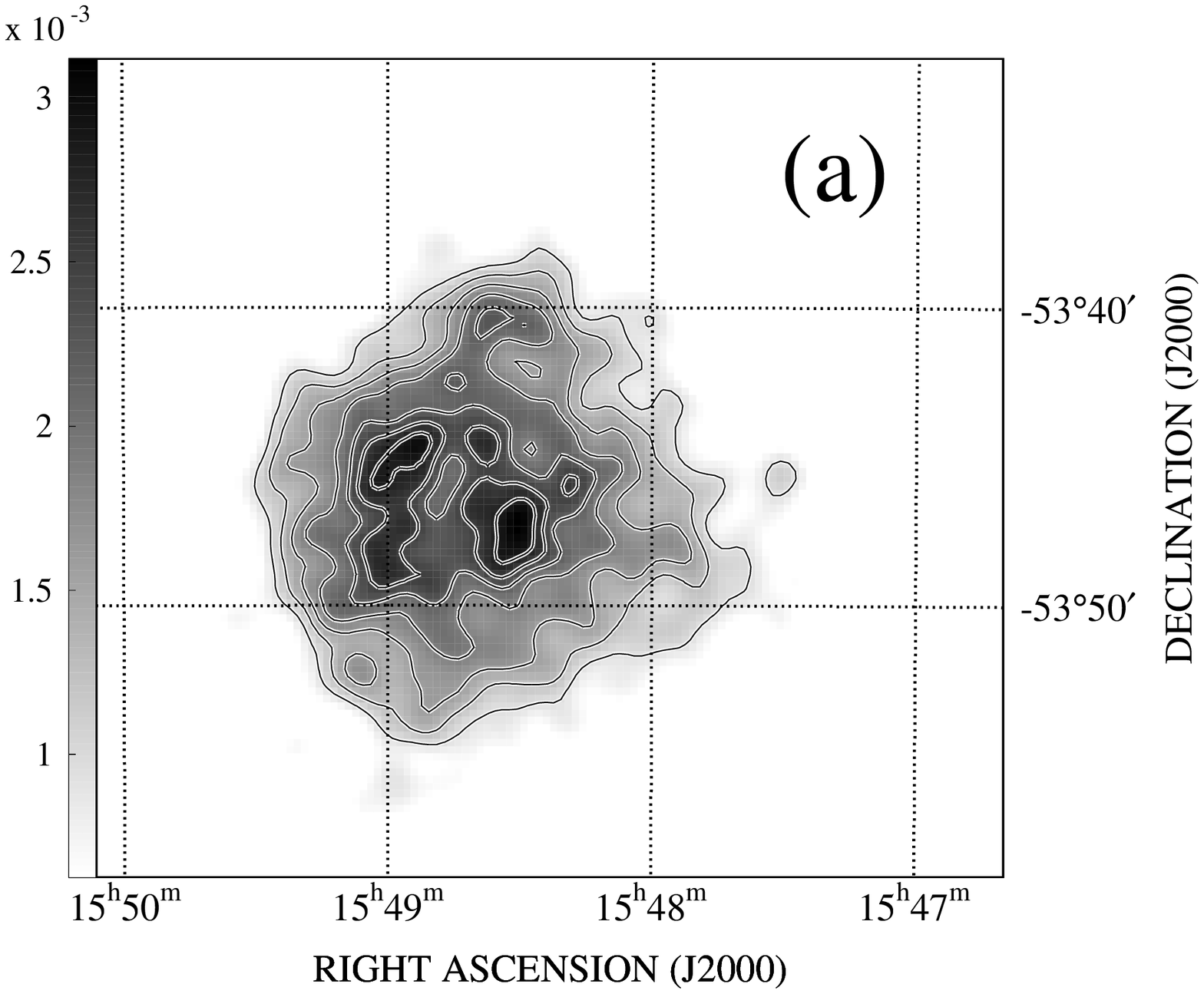}
\FigureFile(85mm,85mm){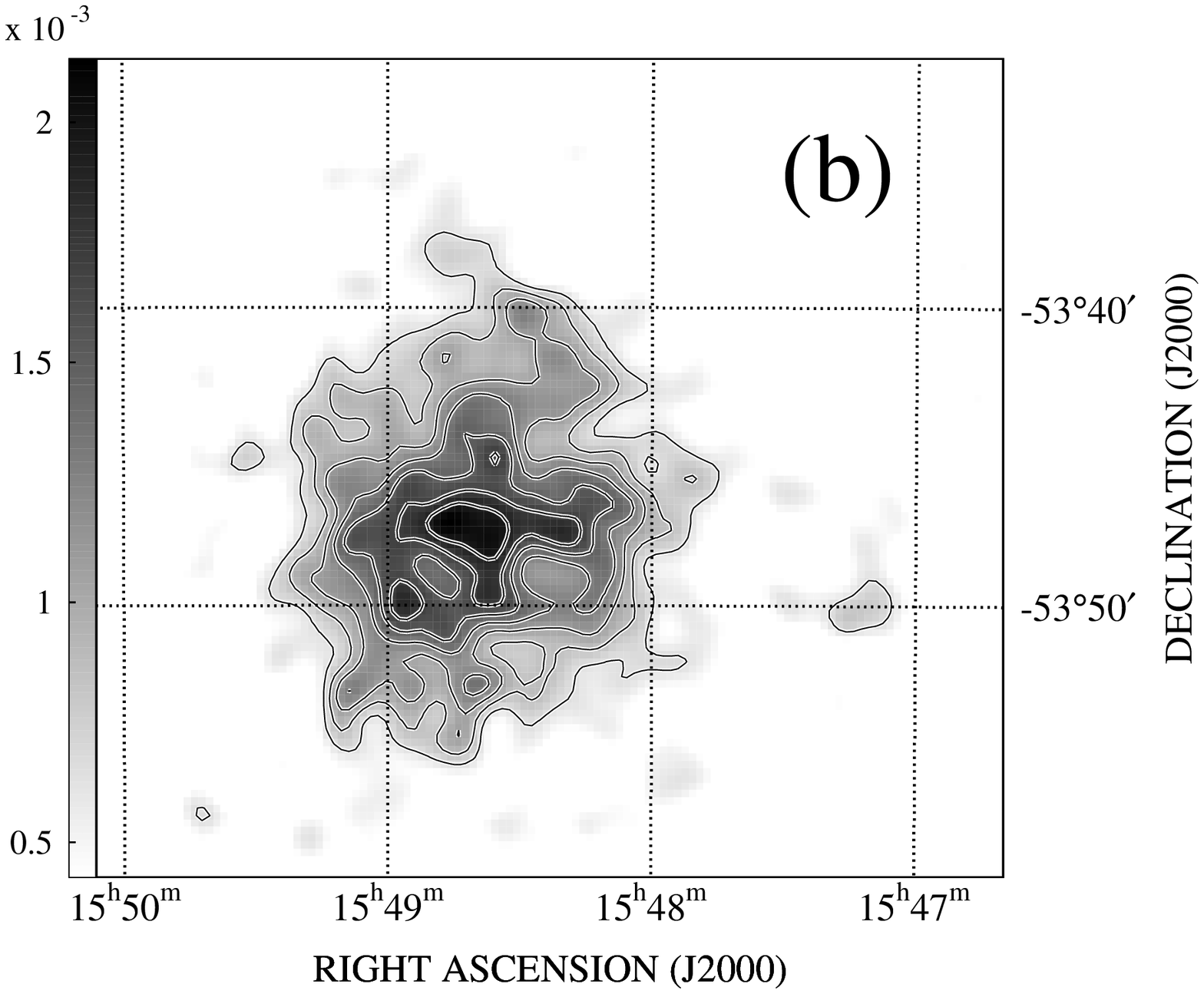}
 \caption{Same as for
 figure~\ref{fig:kes27_gismost} for the 0.7--2\,keV band (a) and for
 the 2--10\,keV band (b).  The contours are linearly spaced X-ray intensity.} 
\label{fig:kes27_gisimage}
\end{figure}

Figure~\ref{fig:kes27_gisimage} shows GIS images for two energy
bands: the low-energy (0.7--2\,keV) band and the high-energy (2--10\,keV)
band.  Comparing the low-energy band image with that obtained by the
PSPC (Seward et al.\ 1996), we noticed that the ASCA image shows
no rim brightening in the eastern rim.  The eastern rim was detected
with the Einstein IPC with relatively weak intensity (Seward
1990).  Considering the difference in the energy bands of these detectors,
the emission from the eastern rim must be soft in the spectrum.

Seward et al.\ (1996) reported several point-like X-ray sources detected
by the PSPC inside the remnant.  Comparing the image obtained by the
HRI, they found that these sources show soft emission.  Due to the
spatial resolution of ASCA, we found no point-like feature
corresponding to them.  The ASCA image shows a centrally peaked
structure that does not coincide with the point source reported.

\begin{figure}
\FigureFile(85mm,85mm){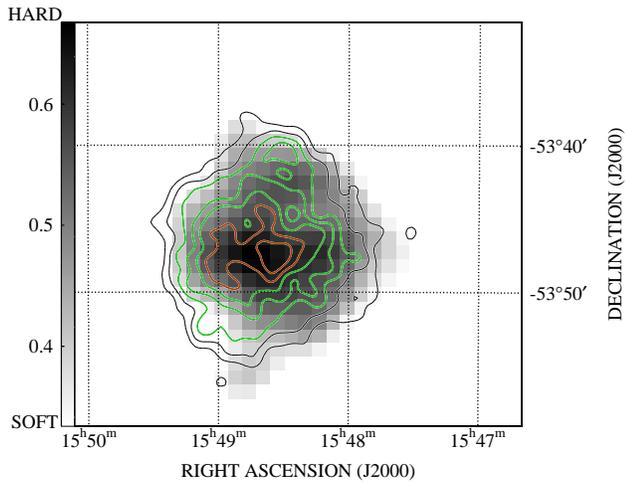}
\caption{GIS band ratio map (2--10\,keV/0.7--2\,keV).  The contours
shows the X-ray intensity (0.7--10\,keV), as shown in
 figure~\ref{fig:kes27_gismost}: red is the strongest, followed by
 green and black.}
 \label{fig:kes27_gishard} 
\end{figure}

We generated the GIS band ratio map by dividing the 2--10\,keV image by
the 0.7--2\,keV image, as shown in figure~\ref{fig:kes27_gishard}.  It
shows that the central part shows a harder spectrum than that of the outer
region.  The hardest region statistically coincides with the X-ray
brightest region, which might suggest the existence of a point source in
the center.

\subsection{Spectral Analysis}\label{subsec:kes27_spec}

\begin{figure}
\FigureFile(85mm,85mm){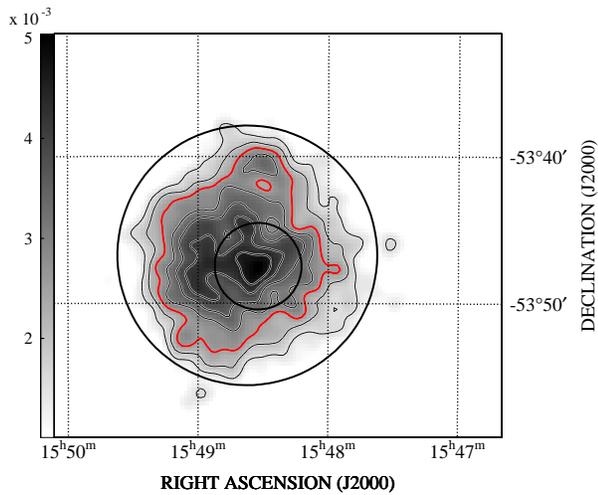}
\caption{ASCA GIS image in the energy range 0.7--10\,keV.  The
gray-scale bar is in units of count s$^{-1}$ arcmin$^{-2}$.  The
contour values are linearly spaced from 0.3 to 0.9 of the peak surface
brightness.  The circles in the solid line indicate the regions used for
the spectral analysis. The half-maximum intensity is also drawn by the
 thick red solid contour.}
\label{fig:kes27_gis0710}
\end{figure}

We extracted the GIS and SIS spectra from the circular region with a
diameter of 18$^{\prime}$, which is shown by the outermost circle in the 
thick
solid line in figure~\ref{fig:kes27_gis0710}, which surely covers the
entire X-ray emission observed.  Since Kes\,27 is located near the
Galactic plane, ($l$, $b$) = (\timeform{327D.4},\timeform{+0D.4}), the
Galactic ridge emission should be taken into account.  Therefore, we
extracted the annular region around the source as the background
spectra.  We selected the annular region of inner and outer diameters of
24$^{\prime}$ and 30$^{\prime}$ as the background for the GIS data.
Similarly, we selected the region of the SIS fov outside a circle of
22$^{\prime}$ diameter as background for the SIS data.  In the
subtraction process, we took into account the vignetting effect of the
X-ray optics.  After background subtraction, we found that there is
no emission above 4\,keV.  We, therefore, employed the spectra up to
4\,keV for a spectral analysis. 

\begin{figure}
\FigureFile(85mm,85mm){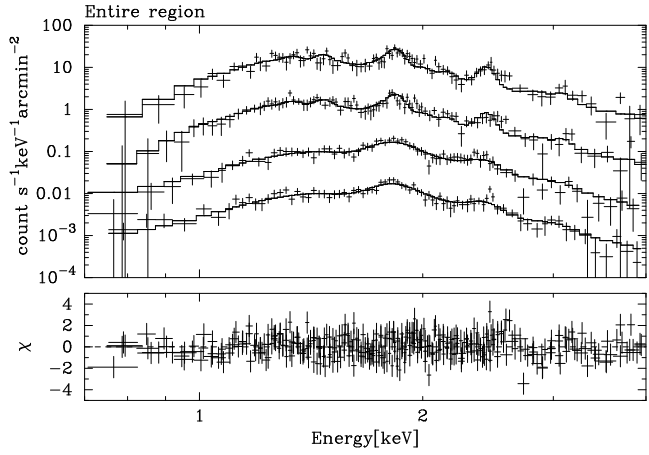}
\caption{Spectra from the entire region of Kes\,27.  Best-fit VMEKAL
models are shown by the solid lines. The spectra correspond to SIS~0 ($\times$ 100), SIS~1 ($\times$ 10), GIS~2, and GIS~3 ($\times$ 0.1) from top to bottom.  The lower panel shows the residuals.}
\label{fig:kes27_spec_entire}
\end{figure}

We performed a simultaneous fit of the model using the GIS and SIS
data.  From the spectra in figure~\ref{fig:kes27_spec_entire}, the
emission lines of Si and S are clearly seen, which indicates that the
X-ray emission comes from a thin thermal plasma in origin.  We employed
the VMEKAL model in {\sc ftool~5.1} where the free parameters were the
interstellar absorption feature, $N_{\rm H}$, electron temperature,
$kT_{\rm e}$, abundances of Mg, Si, S, Ar, and Fe.  The other metal
abundances were fixed to the solar values.  We obtained a statistically
good fit, the results of which are listed in
table~\ref{tbl:kes27_fitpar}, of the entire region. The best-fit curves
are also shown in figure~\ref{fig:kes27_spec_entire}.  We should note
that the obtained abundances are statistically consistent with those of
the solar values. In the spectral fitting, we need not require to employ
a model with the non-equilibrium ionization (NEI) condition.  The
ionization parameter, $\tau$, the product of the electron density and
the elapsed time after the shock  heating, is longer than
10$^{12}$\,cm$^{-3}$s, which indicates that the plasma almost reaches
the collisional ionization equilibrium (CIE) condition.  Using the
plasma density, the value of $\tau$ gives us an age estimate of longer
than $8 \times 10^{4}$ yr, suggesting that the SNR is in the radiative stage.

\begin{figure}
\FigureFile(85mm,85mm){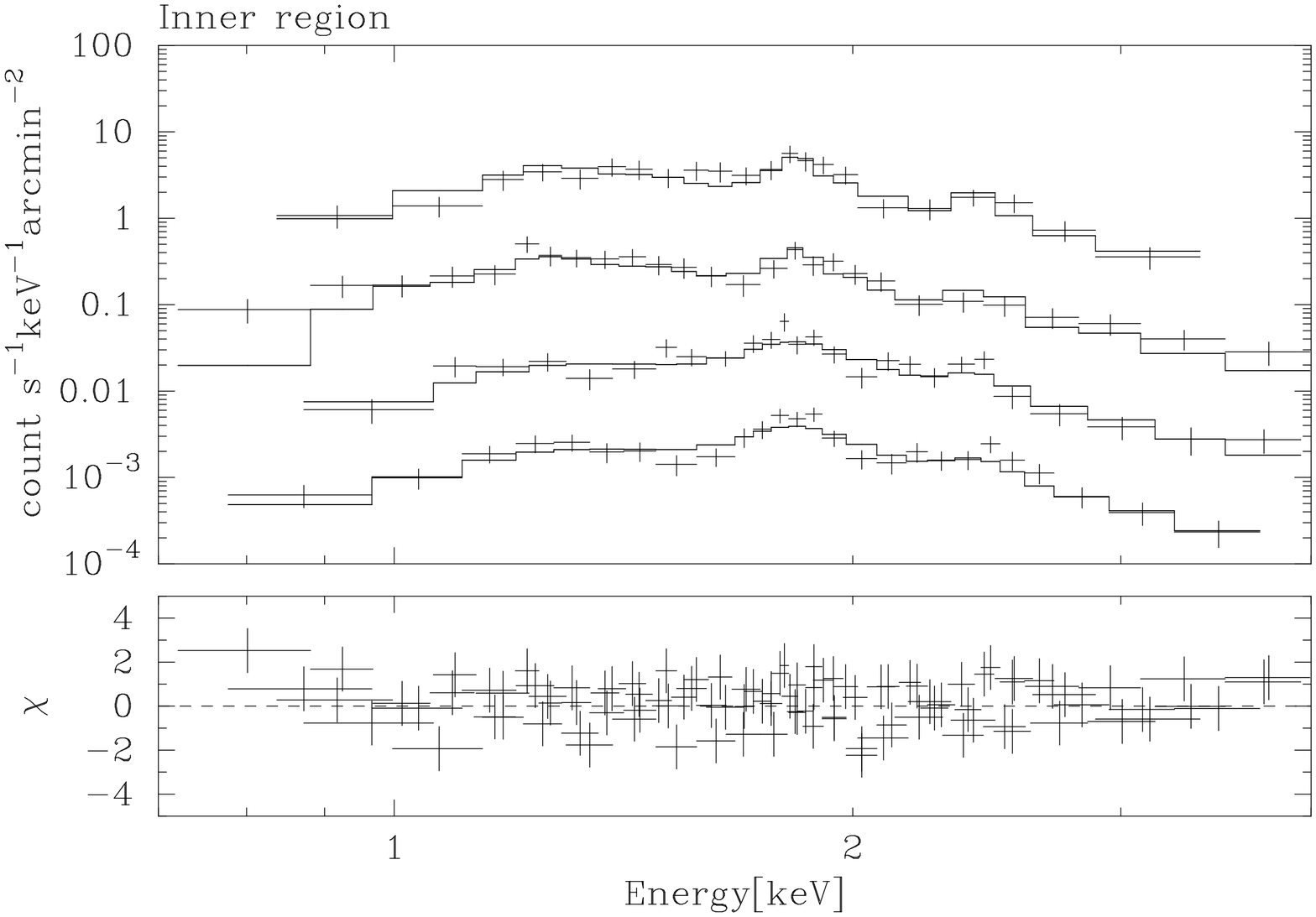}
\FigureFile(85mm,85mm){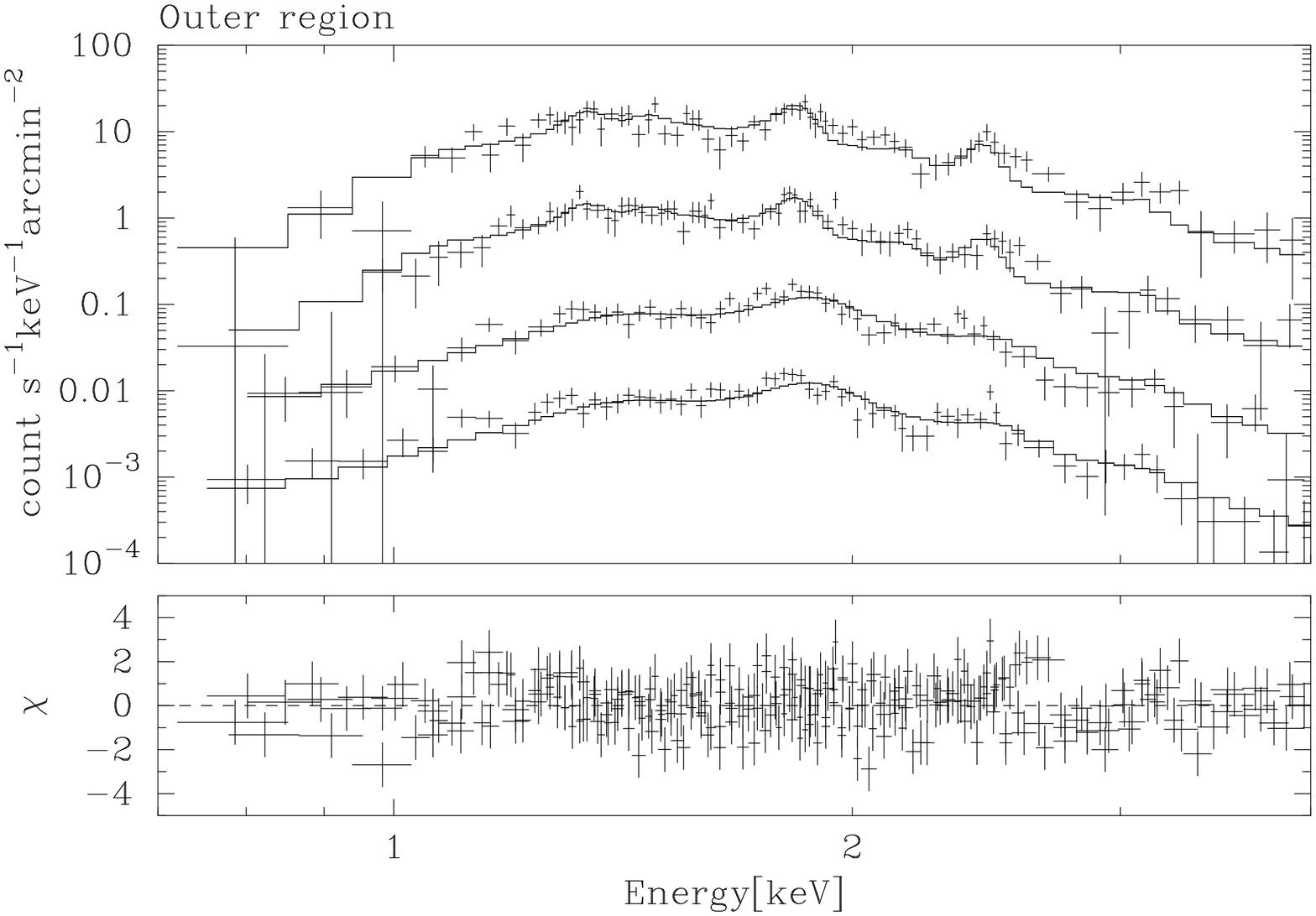}
\caption{Same as figure~\ref{fig:kes27_spec_entire}, but for the inner
region (left) and the outer region (right).}
\label{fig:kes27_speceach}
\end{figure}

We then tried to search for a spectral variation across the
remnant. Since the X-ray band ratio map shows a point symmetric structure, we
divided the image into two regions: an inner region and an outer
region.  The inner region has a diameter of 6$^{\prime}$ centered on the
intensity peak in figure~\ref{fig:kes27_gis0710}.   The outer region has 
an 18$^{\prime}$ diameter.  Figure~\ref{fig:kes27_speceach}
shows the spectra obtained from the two regions.  We also performed a model
fitting by employing the same model to that employed for the entire region.
The best-fit parameters are given in table~\ref{tbl:kes27_fitpar} as
well as the best-fit curves in figure~\ref{fig:kes27_speceach}.  We can
notice that most of them show statistically the same value, with the
exception of $kT_{\rm e}$.  We can say that there is a temperature decrease
from the inner region towards the outer region that causes the harder
spectra in the  center.

As we mentioned before, the X-ray brightest region does not coincide
with the point source detected by ROSAT.  If there is a point
source inside the SNR, it usually shows a hard spectrum and a possible
pulsation.  Seward et al.\ (1996) noticed the possibility that enhanced
central emission could be due to a synchrotron nebula. Our results do
not require a power-law type component, like that expected for a
possible point source inside the SNR.  We picked up photons in the
2--10\,keV range of the GIS from the inner region.  The total number of
photons was 381 with a time resolution of 62.5\,ms.  A Fourier analysis
showed no prominent pulsation component between 0.125--2000\,s.  Due to
the poor statistics, we could not obtain a meaningful upper limit for
the pulsation.  The origin of the central emission is still unknown.

\section{Discussion}

\subsection{Properties of Kes\,27}

So far, there is  no reliable distance estimate to Kes\,27.  Seward et
al.\ (1996) assigned a distance of 6.5\,kpc by comparing
the column density with that of the neighboring SNR, RCW\,103
(5$^{\circ}$ away from Kes\,27), whose distance was determined by a
H\emissiontype{I} measurement.  Case and Bhattacharya (1998) used a
refined $\Sigma$--{\it D} relation to derive a distance of 4.0\,kpc.
Here, we adopt a distance of 5\,kpc, an intermediate value of two estimates.

In order to determine the extent of the X-ray emitting plasma, we
measured the solid angle subtended by a half-maximum intensity contour.
In figure~\ref{fig:kes27_gis0710}, the half-maximum intensity contour is
drawn by the curved red line.  The solid angle subtended by this contour
corresponds to a circle with a diameter of \timeform{11'.5}.  Since
Kes\,27 shows a filled-center structure with a point-symmetric shape, we
assume that the X-ray emitting plasma occupies a sphere with diameter
$d$ and a uniform density.  If this is the case, the half-maximum
intensity contour becomes a circle with a diameter of
$\frac{\sqrt{3}}{2}d$.  Therefore, we calculated various parameters
while assuming the diameter of the plasma to be 13$^{\prime}$ with a uniform
density.  We also assumed the electron density, $n_{\rm e}$, to be equal to the
proton density, $n_{\rm H}$, for simplicity.  These are summarized in
table~\ref{tbl:kes27_prop}.

vWe performed a similar image analysis using the ROSAT PSPC. We
found that the solid angle subtended by the half-maximum intensity
contour is equal to a circle with a diameter of 12$^{\prime}$, which is 
almost equal to that by using the ASCA GIS. This can be understood
because Kes\,27 has such a  high $N_{\rm H}$ that the emission is
confined to only the ASCA energy range.

The apparent size in the radio region is $23^{\prime}\times19^{\prime}$
(Whiteoak, Green 1996), which is bigger by 50\% in size than that we
measured in the 
X-ray region.  The density given in table~\ref{tbl:kes27_prop} is that
of the X-ray emitting plasma.  The total mass in the X-ray emitting region
is a part of the mass affected by the SN explosion.  The radio-bright
region outside the X-ray emitting region must have a higher density than
that inside, since it cools down and becomes X-ray dim.  Judging
from the apparent sizes in the radio and X-ray regions, the total mass occupied
outside the X-ray bright region must be by at least one order magnitude
bigger than that in the X-ray bright region.  Therefore, if the ejecta
contain plenty of metal, it must be dissolved into the ISM, resulting in
the solar abundance.  This shows that Kes\,27 is a typical MM SNR.

\subsection{Spectral Variation}

In the radio band, Kes\,27 shows a bright structure along the eastern
rim.  The bright radio feature on one side of the remnant resembles the
radio morphology of SNRs that are known to show an interaction with the
molecular cloud, such as 3C\,391 and W\,44.  In general, a detection of
shock-excited OH maser emission is a strong evidence for an interaction
with the molecular cloud.  The detection of OH maser emission was
reported for 3C\,391 (Frail et al.\ 1996) and W\,44 (Claussen et al.\ 1997
and references therein), while no OH emission has been detected toward
Kes\,27 (Frail et al.\ 1996).  Therefore, Rho and Petre (1998) noted that
there is no evidence of an actual interaction with the molecular cloud
in Kes\,27.

The ASCA observation verified the centrally peaked and thin
thermal X-ray nature for Kes\,27.  In addition, we first found a
spectral variation from the inner region towards the outer region.  It
should be noted that the harder spectrum in the inner region comes not
from the heavier interstellar absorption but from the higher
temperature.  The value of $kT_{\rm e}$ in the inner region is higher by
40\% than that in the outer region, while there is no variation of the
abundances of heavy elements from a statistical point of view.  
The temperature gradient suggests that thermal conduction does
not play an important role. However, our analysis may be an
oversimplification due to showing an one-temperature 
model. Other SNRs having both a radio shell and centrally peaked
X-ray emission with a thermal nature, like Kes\,27, have a uniform
temperature distribution.  For example, no significant $kT_{\rm e}$ variation
has been found in 3C\,400.2 (Yoshita et al.\ 2001), G69.4+1.2 (Yoshita
et al.\ 2000).   

As mentioned in subsection~3.1, there are two radio-bright regions in
Kes\,27: one is in the east rim and the other is in the center.  The
bright-radio features are generally seen in the rim of the SNR where the
matter is compressed by a blast wave.  When the matter is compressed,
the polarization becomes prominent.  The polarization map given by Milne
et al.\ (1989) shows strong polarization in the east rim as well as in
the center where both regions are radio-bright.  However, the east rim
is X-ray dim and the center is X-ray bright.  Therefore, it must be a
projection effect that the bright radio feature in the center coincides
with the X-ray bright region.  Future X-ray observations with a higher
spatial resolution, such as by using Chandra or XMM-Newton, will reveal
the relation between the X-ray intensity peak and the radio emission.  

\section{Conclusion}

We observed Kes\,27, a proto-typical MM SNR, using ASCA GIS
and SIS data.  We confirmed a filled-center structure in the X-ray
region with an angular diameter of 13$^{\prime}$, which is about 50\%
smaller than that of the radio bright region.  The X-ray spectrum can be
well fitted by a CIE (VMEKAL) model with solar abundances.  Comparing
the spectrum from the inner region with that from the outer region, all
of the spectral parameters are consistent with each other, with an
exception of $kT_{\rm e}$. The inner region clearly shows a higher value
of $kT_{\rm e}$ by 40\% than
that in the outer region.  This suggests that the thermal conduction
does not play an important role.  Our analysis using ASCA supports
that Kes\,27 is a proto-typical MM SNR. 

There are two radio-bright regions in Kes\,27: one is in the east rim
and the other is in the center.  The east rim is X-ray dim while the
center is X-ray bright.  By taking into account the radio
polarization, the coincidence between the X-ray and the radio in the
center is found to be a projection effect.  A finer observation in the
future will reveal the detailed structure in the X-ray region as well as any
coincidence between the X-ray and radio regions. 

\bigskip
The authors would like to express their special thanks to the ASCA team.
This research was partially supported by the Grant-in-Aid for Scientific
Research by the Ministry of Education, Culture, Sports, Science and
Technology of Japan (13440062, 13874032).

\newpage

\begin{table*}  
 \begin{center}
 \caption{Spectral-fit parameters for Kes\,27 in various regions.}
 \label{tbl:kes27_fitpar}
  \begin{tabular}{lccc}
   \hline\hline
   Parameter & Entire & Inner & Outer \\ 
\hline
   $N_{\rm H} [10^{22}\,{\rm cm}^{-2}$] \dotfill & 2.4$^{+0.3}_{-0.2}$ &
   2.2$\pm$0.3 & 2.6$^{+0.4}_{-0.3}$ \\
   $kT_{\rm e}$[keV] \dotfill & 0.71$^{+0.04}_{-0.06}$ & 
   0.84$\pm$0.08 & 0.59$^{+0.04}_{-0.06}$ \vspace{0.3cm}\\
   Mg \dotfill & 1.1$^{+0.6}_{-0.3}$ &
   $< 1.0$ & 1.1$^{+0.9}_{-0.4}$ \\
   Si \dotfill & 0.9$\pm$0.2 &
   0.9$^{+0.6}_{-0.3}$ & 0.7$\pm$0.2 \\
   S \dotfill & 1.1$\pm$0.2 &
   1.2$^{+0.6}_{-0.5}$ & 0.9$\pm$0.3 \\
   Ar \dotfill & $< 1.7$ &
   $< 1.8$ & $< 2.1$ \\
   Fe \dotfill & 1.2$^{+1.1}_{-0.5}$ &
   1.5$^{+1.5}_{-0.9}$ & 0.9$^{+1.9}_{-0.5}$ \\
   $\chi^2$/d.o.f. \dotfill & 333/342 &
   102/90 & 339/280 \\
   \hline
  \\

   & & & \\[-15pt] 
  \multicolumn{3}{l}{Note. Other elements are fixted to those of solar values.}
   \\
  \multicolumn{3}{l}{Errors are the 90\% confidence level.} 
  \end{tabular}
 \end{center}
\end{table*}

\newpage

\begin{table*} 
 \begin{center}
  \caption{Physical parameters of Kes\,27.}
  \label{tbl:kes27_prop}
  \begin{tabular}{lc}
   \hline\hline
   Parameter\hspace*{1cm} & Value \\ \hline
   Diameter [pc] \dotfill & 18.9$d_{5}$ \\
   Temperature [keV] \dotfill & 0.71$^{+0.04}_{-0.06}$ \\
   Density [cm$^{-3}$] \dotfill & 0.39$^{+0.03}_{-0.02}$ $d^{-1/2}_{5}$ \\
   Thermal energy [erg] \dotfill & $(1.4\pm 0.1)\times 10^{50}$ $d^{5/2}_{5}$ \\
   Total Mass [$M_{\odot}$] \dotfill & 34$^{+3}_{-2}$ $d^{5/2}_{5}$ \\ \hline
\\
\\[-15pt] 
   \multicolumn{2}{l}{Note. $d_{5}$ is the distance in units of 5\,kpc.} \\
  \end{tabular}
 \end{center}
\end{table*}

\end{document}